\documentclass[11pt, a4paper, twoside]{article}
\input{epsf}
\def\be{\begin{equation}}
\def\ee{\end{equation}}

\def\half{\frac{1}{2}}
\def\third{\frac{1}{3}}

\def\mgap{\qquad \qquad}

\def\ltsim{\lower3pt\hbox{$\, \buildrel < \over \sim \, $}}

\def\similar{\quad {\mathop\sim\limits_{\scriptscriptstyle
 n \to \infty} } \quad}

\def\mone#1{M^{(1)}_{#1}}
\def\mean#1#2{M^{(#1)}_{#2}}
\def\semi#1#2{\overline{M}^{(#1)}_{#2}}

\def\xx{|{\bf X}|^2}
\def\xa{X_a}
\def\xb{X_b}
\def\xc{X_c}
\def\xad{X_a^\dagger}
\def\xbd{X_b^\dagger}
\def\xcd{X_c^\dagger}

\def\lang{\left\langle}
\def\rang{\right\rangle}

\def\oh{\hat{O}}
\def\ddy{\frac{\partial}{\partial y}}

\def\rb{\overline{R}}

\thicklines

\def\plus{ + }

\def\one{\begin{picture}(40,30)
  \put(20,0){\oval(30,60)[t]}
  \end{picture}}
\def\two{\begin{picture}(56,50)
  \put(28,0){\oval(46,90)[t]}
  \put(28,0){\oval(24,55)[t]} 
  \end{picture}}
\def\twoinone{\begin{picture}(70,50)
  \put(35,0){\oval(60,90)[t]}
  \put(22,0){\oval(20,45)[t]}
  \put(48,0){\oval(20,45)[t]}
  \end{picture}}
\def\three{\begin{picture}(70,50)
  \put(35,0){\oval(60,90)[t]}
  \put(35,0){\oval(40,60)[t]}
  \put(35,0){\oval(20,40)[t]} 
  \end{picture}}

\def\bone{\begin{picture}(40,30)(0,-30)
  \put(20,0){\oval(30,60)[b]}
  \end{picture}}
\def\btwo{\begin{picture}(56,50)(0,-30)
  \put(28,0){\oval(46,90)[b]}
  \put(28,0){\oval(24,55)[b]} 
  \end{picture}}
\def\btwoinone{\begin{picture}(70,50)(0,-30)
  \put(35,0){\oval(60,90)[b]}
  \put(22,0){\oval(20,45)[b]}
  \put(48,0){\oval(20,45)[b]}
  \end{picture}}
\def\bthree{\begin{picture}(70,50)(0,-30)
  \put(35,0){\oval(60,90)[b]}
  \put(35,0){\oval(40,60)[b]}
  \put(35,0){\oval(20,40)[b]} 
  \end{picture}}

\def\xline{\thinlines
  \line(0,1){40}
  \thicklines}

\def\xlinelong{\thinlines
  \line(0,1){60}
  \thicklines}

\def\xsplit{\begin{picture}(40,40)
  \put(20,0){\oval(30,60)[t]}
  \put(20,0){\xline}
  \end{picture}}

\def\xsplitlong{\begin{picture}(40,40)
  \put(20,0){\oval(30,60)[t]}
  \put(20,0){\xlinelong}
  \end{picture}}

\def\xsplittwo{\begin{picture}(56,50)
  \put(28,0){\oval(46,90)[t]}
  \put(28,0){\oval(24,55)[t]}
  \put(28,0){\xlinelong}
  \end{picture}}

\def\xsplittwoleft{\begin{picture}(90,50)
  \put(45,0){\oval(80,90)[t]}
  \put(35,0){\oval(24,55)[t]}
  \put(65,0){\xlinelong}
  \end{picture}}

\def\xisleft{\begin{picture}(110,50)
  \put(50,0){\oval(100,90)[t]}
  \put(30,0){\oval(24,55)[t]}
  \put(50,0){\xsplitlong}
  \end{picture}}

\def\xisright{\begin{picture}(110,50)
  \put(50,0){\oval(100,90)[t]}
  \put(70,0){\oval(24,55)[t]}
  \put(10,0){\xsplitlong}
  \end{picture}}

\def\xsplitthree{\begin{picture}(70,55)
  \put(35,0){\oval(60,100)[t]}
  \put(35,0){\oval(40,70)[t]}
  \put(35,0){\oval(20,50)[t]}
  \put(35,0){\xlinelong}
  \end{picture}}

\def\toright{\begin{picture}(40,40)
  \put(20,0){\oval(30,50)[t]}
  \put(0,30){$\rightarrow$}
  \end{picture}\ }

\def\toleft{\begin{picture}(40,40)
  \put(20,0){\oval(30,50)[t]}
  \put(0,30){$\leftarrow$}
  \end{picture}\ }

\def\toboth{\toright\! + \toleft}

\def\tofact{\Bigl( \toboth\! \Bigr)}
\def\twofact{\Bigl( \toright\!\!\toright\! + \toleft\!\!\toleft\! \Bigr)}
\def\threefact{\Bigl(\toright\!\!\toright\!\!\toright\! + \toleft\!\!\toleft
\!\!\toleft\! \Bigr)}

\def\dotline{. \ \xlinelong}

\def\twounder{\left(
\twounderplain
\right)}

\def\twounderplain{\begin{picture}(120,65)
\put(60,0){\oval(120,110)[t]}
\put(10,0){\one \one \ \xlinelong}
\end{picture} 
+
\begin{picture}(120,65)
\put(60,0){\oval(120,110)[t]}
\put(10,0){\xlinelong \one \one}
\end{picture}}

\def\twosplitplain{\begin{picture}(140,65)
\put(70,0){\oval(140,110)[t]}
\put(10,0){\one \one \xsplitlong}
\end{picture} 
+
\begin{picture}(140,65)
\put(70,0){\oval(140,110)[t]}
\put(10,0){\xsplitlong \one \one}
\end{picture}}

\def\threeunder{\left(
\begin{picture}(160,65)
\put(80,0){\oval(160,110)[t]}
\put(10,0){\one \one \one \ \xlinelong}
\end{picture} 
+
\begin{picture}(160,65)
\put(80,0){\oval(160,110)[t]}
\put(10,0){\xlinelong \one \one \one}
\end{picture}
\right)
}

\begin{document}

\hfill    NBI-HE-98-18

\hfill July 1998

\begin{center}
\vspace{20pt}
{\Large \bf A Diagrammatic Approach to the Meander Problem}

\vspace{30pt}

{\sl M. G. Harris}\footnote{E-mail: Martin.Harris@nbi.dk}

\vspace{20pt}
 The Niels Bohr Institute,\\
Blegdamsvej 17, DK-2100 Copenhagen \O , Denmark.\\

\vspace{12pt}
  
\end{center}
\vspace{12pt}

\vfill

\begin{center}
{\bf Abstract}
\end{center}

\vspace{6pt}

\noindent

The meander problem is a combinatorial problem which provides a toy
model of the compact folding of polymer chains. In this paper we study
various questions relating to the enumeration of meander diagrams,
using diagrammatical methods. By studying the problem of folding tree
graphs, we derive a lower bound on the exponential behaviour of the
number of connected meander diagrams. A different diagrammatical
method, based on a non-commutative algebra, provides an
approximate calculation of the behaviour of the generating functions
for both meander and semi-meander diagrams.

\bigskip



\bigskip


\noindent

\bigskip

\vfill

\newpage


\section{Introduction}
\label{sec:intro}

The problem of enumerating meander and semi-meander diagrams has
appeared in many diverse areas of mathematics~\cite{Arn88,KS91} and
computer science~\cite{HMRT,LZ}. In the case of semi-meanders it can be
reformulated as an enumeration of the foldings of a strip of postage
stamps~\cite{Touch,FGG95}, which is equivalent to a model of compact
foldings of a polymer. Thus the semi-meanders serve as a toy model of
the statistical mechanics of such polymer chains.

The generating function for meanders can be written as an Hermitian
matrix model or as a supersymmetric matrix
model~\cite{Mak95,MkCh96}. Alternatively combinatorial methods can be
used~\cite{Touch,FGG95}, one can represent the problem in terms of
non-commuting variables~\cite{MkCh96} or using a Temperley-Lieb
algebra~\cite{FGGTL,Fran96,Fran97}. 
Despite all of these different
approaches and the initial apparent simplicity of the problem, the
task of counting the meander and semi-meander diagrams has remained
largely unsolved.

In this paper we apply diagrammatical methods to the study of
meanders. In section~\ref{sec:def} the problem is defined and some tables
of meander numbers are given. Then in section~\ref{sec:three} we
calculate bounds on the number of connected meanders, in particular
the lower bound is improved by mapping the meanders to folded tree
graphs. Section~\ref{sec:diagram} uses a diagrammatic representation
of a non-commutative algebra to calculate approximate generating
functions for meanders and semi-meanders. The nature of the 
phase transition for semi-meanders is also considered in this section.
Finally we give our conclusions in section~\ref{sec:concl} of the paper.

\section{Definition of the meander numbers}
\label{sec:def}

\subsection{Meanders}
\begin{figure}[htb]
\begin{center}
\mbox{
\epsfysize2cm
\epsffile{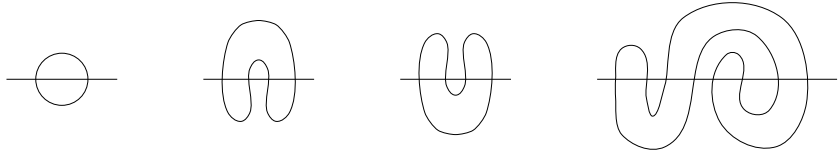}}
\caption{Some simple meander diagrams.}
\label{fig:meander}
\end{center}
\end{figure}

Suppose that we have an infinite line (or ``river'') on a planar
surface. Then a connected meander of order $n$ is defined to be a
non-self-intersecting connected loop (or ``road'') which crosses the
river $2n$ times (the crossing points are referred to as ``bridges'').
Two meanders are considered to be equivalent if it is possible to
smoothly deform one meander into the other without changing the number
of bridges during the process. The total number of inequivalent
connected meanders of order $n$ is denoted $\mone {n}$, and we
can define a generating function, $M(x)$ for connected meanders,
\be
M(x)= \sum_{n=1}^{\infty} \mone n \ x^n .
\ee
The first few meander numbers are $\mone 1=1$, $\mone 2=2$ and $\mone 3=8$.
Figure~\ref{fig:meander} shows some simple meander diagrams.

\begin{figure}[ht]
\begin{center}
\mbox{
\epsfysize2cm
\epsffile{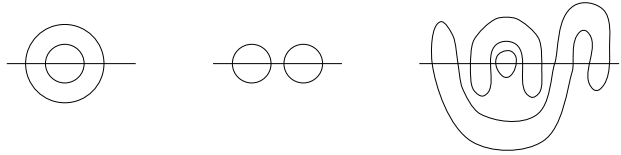}}
\caption{Some simple disconnected meander diagrams.}
\label{fig:dismean}
\end{center}
\end{figure}

The definition can be extended to include cases in which the road
consists of $k$ disconnected parts, giving a number of meanders of
order $n$ denoted by $\mean kn$. This leads us to define a more general
generating function,
\be
M(x,m) = \sum_{n=1}^{\infty} \sum_{k=1}^{n} \mean kn \ x^n m^k.
\ee
Table~\ref{tab:numbers} gives $\mean kn$ for small values of $k$ and
$n$ (these numbers are taken from~\cite{FGG95}). Some examples of
disconnected meanders are shown in figure~\ref{fig:dismean}.

\begin{table}[htb]
\begin{center}
\begin{tabular}{|l||rrrrrrr|} \hline
$k$&$n=$ 1 & 2 & 3 & 4 & 5 & 6 & 7\\ \hline
1 & 1 & 2 & 8 & 42 & 262 & 1828 & 13820\\ 
2 & & 2 & 12 & 84 & 640 & 5236 & 45164\\
3 & & & 5 & 56 & 580 & 5894 &60312\\
4 & & & & 14 & 240 & 3344 &42840\\ 
5 & & & & & 42 & 990 &17472\\
6 & & & & & & 132 &4004\\ 
7 & & & & & & & 429 \\ \hline
\end{tabular}
\caption{Meander numbers, $\mean kn$ (from ref.~\cite{FGG95}). }
\label{tab:numbers}
\end{center}
\end{table}

\subsection{Semi-meanders}
\begin{figure}[ht]
\begin{center}
\mbox{
\epsfysize2cm
\epsffile{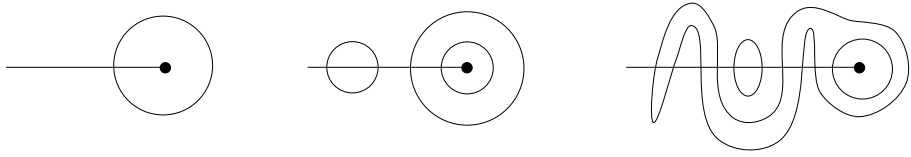}}
\caption{Some semi-meander diagrams. The source is marked with a dot.}
\label{fig:semi}
\end{center}
\end{figure}

Suppose that instead of having an infinite river we have only a
semi-infinite river. The end of the river is often referred to as the
``source''. In this case the road can wrap around the source of the
river and the corresponding diagrams are called ``semi-meanders''. As
before two semi-meanders are equivalent if one can be smoothly
deformed into the other, without changing the number of
bridges. Figure~\ref{fig:semi} shows some examples of semi-meanders.

Let us denote the number of inequivalent semi-meanders with $n$
bridges and a road consisting of $k$ connected parts, as $\semi
kn$. Then the corresponding generating function is
\be
\label{eqn:defsemi}
S(c,m)= \sum_{n=1}^\infty \sum_{k=1}^n \semi kn c^n m^k.
\ee
Note that the meanders can only have an even number of bridges,
whereas semi-meanders can have both even and odd numbers of bridges. This is
the reason for the slightly different definitions of the generating
functions and one should define $x\equiv c^2$. Table~\ref{tab:semi} gives
the numbers for the simplest semi-meanders.

\begin{table}[htb]
\begin{center}
\begin{tabular}{|l||rrrrrrr|} \hline
$k$&$n=$ 1 & 2 & 3 & 4 & 5 & 6 & 7\\ \hline
1 & 1 & 1 & 2 & 4 & 10 & 24 & 66\\ 
2 & & 1 & 2 & 6 & 16 & 48 & 140 \\
3 & & & 1 & 3 & 11 & 37 & 126 \\
4 & & & & 1 & 4 & 17 &66 \\
5 & & & & & 1 & 5 & 24 \\
6 & & & & & & 1 & 6 \\
7 & & & & & & & 1 \\ \hline
\end{tabular}
\caption{Semi-meander numbers, $\semi kn$ (from ref.~\cite{FGG95}). }
\label{tab:semi}
\end{center}
\end{table}

\section{Connected meanders}
\label{sec:three}

\subsection{Bounds on the number of meanders}
\label{sec:bound}

\begin{figure}[ht]
\begin{center}
\mbox{
\epsfysize1.3cm
\epsffile{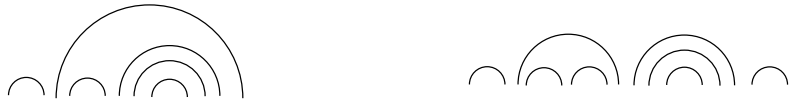}}
\caption{Two example arch diagrams.}
\label{fig:arch}
\end{center}
\end{figure}

\begin{figure}[ht]
\begin{center}
\mbox{
\epsfysize2.5cm
\epsffile{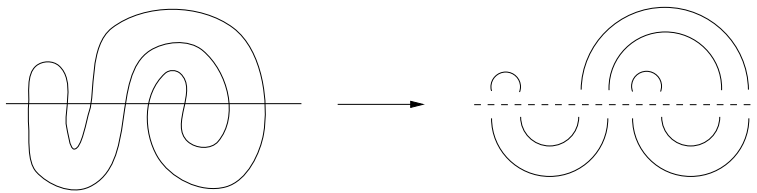}}
\caption{A meander cut into two sets of arches.}
\label{fig:cut}
\end{center}
\end{figure}

In this section we will derive some bounds on the number of inequivalent
connected
meanders (from now on it should be understood that we are referring
always to inequivalent meanders).
The standard way of doing this is to consider a meander as
being formed from two arch diagrams (see figure~\ref{fig:arch} for
some examples of arch diagrams). That
is, if we take a meander diagram and cut it in two along the river,
then this gives one arch diagram above the river and one below it (see
figure~\ref{fig:cut}).

An arch diagram of order $n$ consists of $2n$ points arranged in a
line, which are connected up in pairs by drawing $n$ arches above the
points. The arches are not allowed to intersect each other and each
point has exactly one arch joined to it. The number of inequivalent
order $n$ arch configurations is denoted $A_n$, and the generating
function for arches is defined as
\be
A(c) = \sum_{n=0}^\infty A_n c^n, \mgap A_0 \equiv 1.
\ee
The empty arch diagram $A_0$ has the value $1$.
A non-empty arch diagram can be decomposed into two arch
diagrams separated by the arch connected to the rightmost point. This
is illustrated diagrammatically as
\setlength{\unitlength}{0.1mm}
\begin{picture}(285,50)
\thicklines
\put(230,0){\oval(80,80)[t]}
\put(0,0){\makebox(0,0)[bl]{$A=1 + A \ \ \ A \ \ $}}
\end{picture}.
That is, $A(c)$ satisfies the equation 
\be
\label{eqn:arelation}
A=1+c A^2.
\ee
Hence
\be
\label{eqn:adef}
A(c) = \frac{1}{2c} \left( 1- \sqrt{1-4c}\right)
\ee
and 
\be
A_n=\frac{(2n)!}{n! (n+1)!} \similar  \frac{4^n}{n^{3/2}}.
\ee
In fact $A_n$ is equal to $c_n$,
the Catalan number of order $n$.

Since gluing together two arch configurations of order $n$ gives a
(possibly disconnected) meander of order $n$ we have
\be
\mone n \le \sum_{k=1}^n \mean kn = {A_n}^2.
\ee
In fact, given an arch diagram on top, we can always find an arch
configuration on the bottom that creates a connected meander (using
the algorithm given in fig.~21 of ref.~\cite{FGG95}), which means that
$A_n \le \mone n$.

Consider the asymptotics of $\mone n$ for a large number of bridges,
\be
\label{eqn:masymp}
\mone n \similar  {\it const.} \ \frac{R^{2n}}{n^\alpha},
\ee
where $R$ gives the exponential behaviour and $\alpha$ is a configuration
exponent.
Then
\be
A_n \le \mone n \le {A_n}^2
\ee
implies that 
\be
2 \le R \le 4.
\ee

\subsection{Improved lower bound}
The contents of the previous section are all well-known. In this
section we would like to improve the lower bound on $R$. Instead of
considering arch diagrams we will rewrite the connected meanders as
tree diagrams.

\begin{figure}[ht]
\begin{center}
\mbox{
\epsfysize3cm
\epsffile{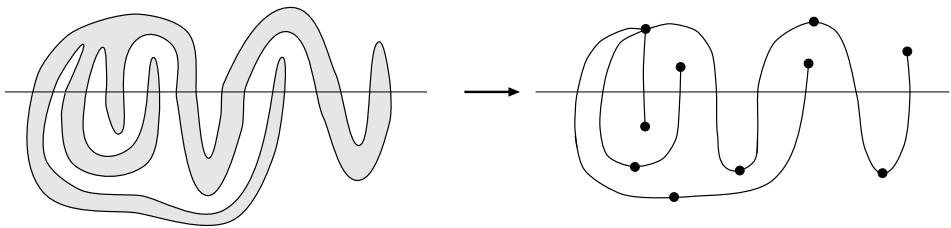}}
\caption{Converting a connected meander, with the inside shaded,
to a (folded) tree diagram.}
\label{fig:tree}
\end{center}
\end{figure}

The road of a connected meander divides the plane into
two parts: an inside and an outside. Smoothly deforming the road in
such a way as to reduce the inside area towards zero yields, in the
limit, a connected tree graph (see fig.~\ref{fig:tree}). Given such a
folded tree graph we can easily reconstruct the corresponding meander.
Note that applying this procedure to a disconnected meander would
yield a graph that is disconnected and/or contains loops. The tree
graphs corresponding to connected meanders can have vertices of
arbitrary coordination number, and the mapping procedure is defined
such that adjacent vertices on the tree occur above and below the
river in an alternating fashion. There is a one-to-one correspondence
between such tree graphs and the set of connected meanders. By putting
bounds on the number of such trees we will improve the lower bound on
$R$ discussed in the previous section.

\begin{figure}[ht]
\begin{center}
\mbox{
\epsfysize3cm
\epsffile{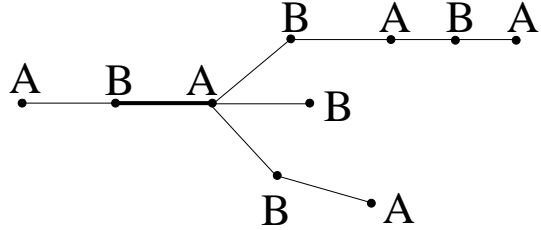}}
\caption{An unfolded tree diagram corresponding to the folded tree in
fig.~\ref{fig:tree}.}
\label{fig:unfold}
\end{center}
\end{figure}

It is convenient to label the vertices of each tree with either an 'A'
or 'B', signifying that they are above or below the river
respectively. The left-most link in the tree will be marked by drawing
it thicker. Each tree can then be unfolded as shown in
figure~\ref{fig:unfold}.

\begin{figure}[htb]
\begin{center}
\mbox{
\epsfysize4cm
\epsffile{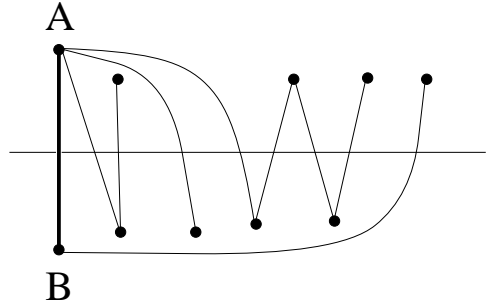}}
\caption{Another more regular folding of the tree diagram.}
\label{fig:refold}
\end{center}
\end{figure}

 In general an unfolded tree has many different ways of folding it,
subject to the constraints that when folded the 'A' vertices are above
the line and the 'B' vertices are below it, and that the marked link is
leftmost. For example, figure~\ref{fig:unfold} can be refolded as in
figure~\ref{fig:refold}. Each unfolded tree has at least one possible folding
and usually has many more. Thus each unfolded tree represents a whole
class of connected meander diagrams.
By counting the number of unfolded trees we will generate a lower
bound on the number of connected meanders.

\begin{figure}[htb]
\begin{center}
\mbox{
\epsfysize1.65cm
\epsffile{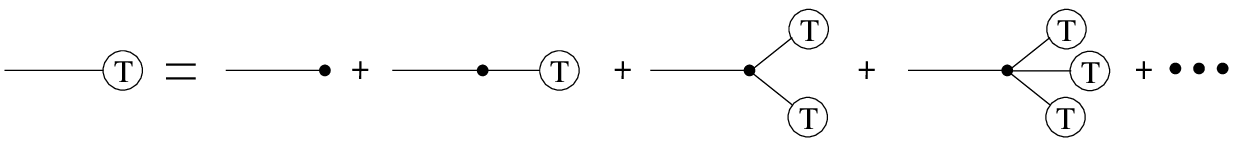}}
\caption{Relation which determines $T(x)$, the generating function for
rooted trees with arbitrary coordination numbers.}
\label{fig:treegen}
\end{center}
\end{figure}

A rooted tree with arbitrary coordination numbers can be generated
using the relation illustrated in fig.~\ref{fig:treegen}. That is, if
each link in the tree is weighted with $x$, then the generating
function for trees, $T(x)$, satisfies
\be
T(x)=x(1+T+T^2+T^3+\cdots)=x/(1-T)
\ee
and hence
\be
T(x)=\half \left(1- \sqrt{1-4x} \right).
\ee

\begin{figure}[tb]
\begin{center}
\mbox{
\epsfysize2.4cm
\epsffile{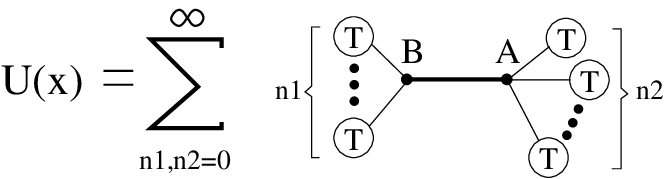}}
\caption{Relation which determines $U(x)$, the generating function for
unfolded trees.}
\label{fig:ugen}
\end{center}
\end{figure}

Unfolded trees have a generating function $U(x)$, which from
fig.~\ref{fig:ugen} is given by
\be
U(x)=\frac{x}{(1-T)^2}.
\ee
Hence
\be
U(x)=\frac{T^2}{x}=\frac{1}{2x} \left(1- \sqrt{1-4x} \right) -1 = A(x)-1,
\ee
with $A(x)$ defined through equation (\ref{eqn:adef}).
So we have
\be
U(x)= \sum_{n=1}^\infty A_n x^n.
\ee
However, as mentioned earlier, the number of unfolded trees (with $n$
links) gives a lower bound on the number of connected meanders (with
$2n$ bridges), that is, $A_n\le \mone n$ and $R \ge 2$.
This reproduces the result in section~\ref{sec:bound}. 
Note that since each link of the tree graph is equivalent to two bridges
in the corresponding connected meander (fig.~\ref{fig:tree}), the
factor for links in the tree had to be $x$ for consistency with the
definitions in section~\ref{sec:def}.

The representation in terms of trees allows us to improve on the bound
by counting more of the possible foldings for each unfolded tree.
Suppose that  the generating function for partially folded trees is
denoted by $\Theta(x)$,
\be
\Theta(x) \equiv \sum_{n=1}^\infty \Theta_n x^n,
\ee
where by partially folded trees, it is meant that only some of the
possible foldings are included for each unfolded tree.
That is, we are summing over all the trees in $U(x)$, but only some of
the foldings. Thus we will have $A_n \le \Theta_n \le \mone n$.
Let us now define $\Theta(x)$ and hence the subset of the possible
foldings that are included in the partial folding. First, the
relation in fig.~\ref{fig:ugen} is rewritten to include more foldings
\begin{figure}[htb]
\begin{center}
\mbox{
\epsfysize3cm
\epsffile{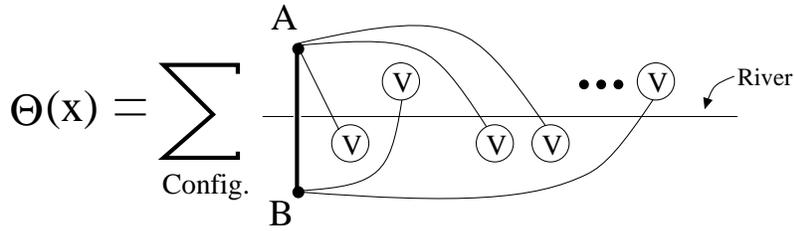}}
\caption{Relation which determines $\Theta(x)$, the generating function for
partially folded trees.}
\label{fig:theta}
\end{center}
\end{figure}
(see fig.~\ref{fig:theta}). The rooted trees $T(x)$ have been replaced
by $V(x)$, which is more folded that $T$ and is defined below.
The sum is over all possible ways of interleaving the arbitrary number
of $V$-trees. This relation gives
\be
\label{eqn:theta}
\Theta(x) = \frac{x}{1-2V},
\ee
where the $x$ comes from the marked link. The $(1-2V)^{-1}$ factor is
due to the sum over all possible numbers of $V$-trees, with each tree
being connected to one of 2 vertices ('A' or 'B').

\begin{figure}[t]
\begin{center}
\mbox{
\epsfysize2.0cm
\epsffile{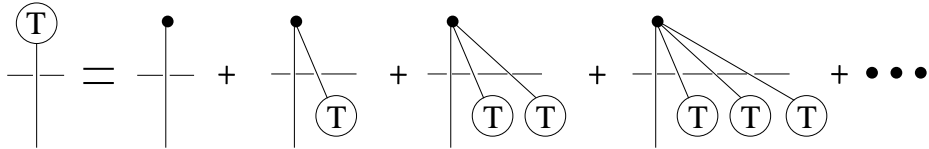}}
\caption{The relation determining $T(x)$, which has been rewritten.}
\label{fig:newtgen}
\end{center}
\end{figure}

\begin{figure}[htb]
\begin{center}
\mbox{
\epsfysize1.6cm
\epsffile{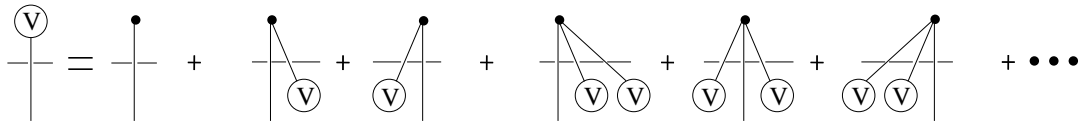}}
\caption{Relation which determines $V(x)$.}
\label{fig:vgen}
\end{center}
\end{figure}

\begin{figure}[htb]
\begin{center}
\mbox{
\epsfysize3.7cm
\epsffile{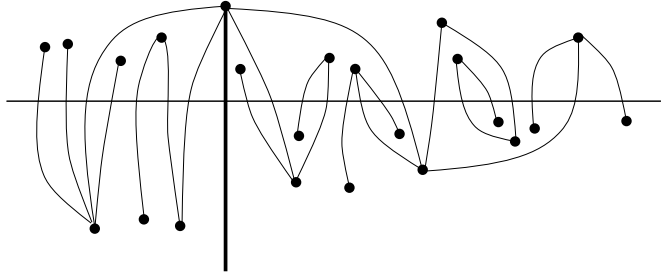}}
\caption{A typical $V$-tree contained in $V(x)$.}
\label{fig:vcomplex}
\end{center}
\end{figure}

Now consider redrawing the relation in fig.~\ref{fig:treegen} as shown
in fig.~\ref{fig:newtgen}. 
Then we see that the generating function
for $T$ only folds in one direction, but that more foldings can be
included if we fold to both left and right. Thus for $V(x)$ the
relation in fig.~\ref{fig:vgen} will be used
instead. Figure~\ref{fig:vcomplex} shows a typical $V$-tree.

Thus we have
\be
\label{eqn:vgen}
V(x) = x\left( 1 + 2V + 3V^2 + 4V^3 + \cdots \right) = \frac{x}{(1-V)^2}.
\ee

\begin{figure}[tb]
\begin{center}
\mbox{
\epsfysize3.4cm
\epsffile{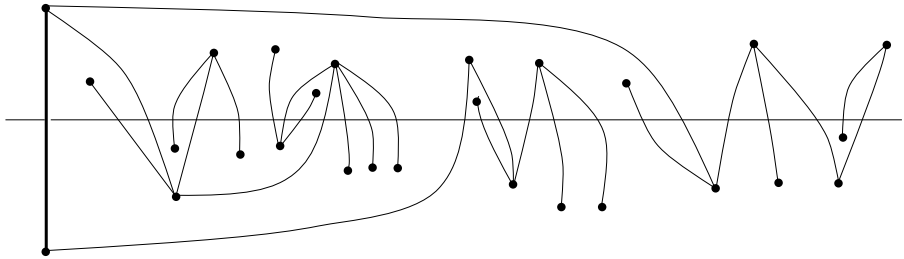}}
\caption{Example of a diagram included in $\Theta(x)$.}
\label{fig:complex}
\end{center}
\end{figure}

Diagrams such as that in fig.~\ref{fig:complex} are generated when
this generating function for $V(x)$ is inserted in to (\ref{eqn:theta}).
Thus $\Theta(x)$ will contain a large number, but my no means all, of the
possible folded tree diagrams. As $x$ is increased from zero it reaches some
critical value $x_c$ at which $\Theta(x)$ is non-analytic. This is
caused by the critical behaviour of $V(x)$ due to $\frac{dV}{dx}$
diverging as $x\to x_c$. Now,
\be
\frac{dV}{dx} = \frac{1}{(1-4V + 3V^2)},
\ee
so that $V(x_c)=\third$ and $x_c=\frac{4}{27}$. Thus
\be
\Theta_n \sim \left(\frac{27}{4}\right)^n
\ee
and hence $R \ge \frac{3 \sqrt{3}}{2} \approx 2.598$. This improves
the bound given previously.

\subsection{Further improvement to lower bound}
\label{sec:further}
\begin{figure}[htb]
\begin{center}
\mbox{
\epsfysize2.2cm
\epsffile{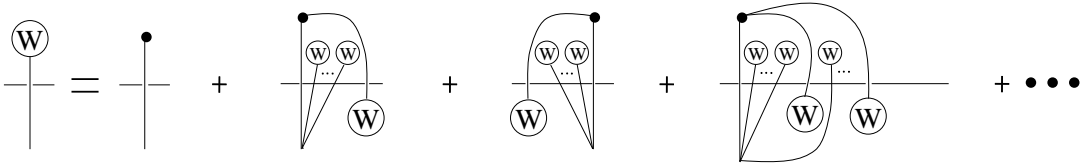}}
\caption{Relation which determines $W(x)$, where we are summing over
arbitrary numbers of $W$-trees inserted below each branch.}
\label{fig:wgen}
\end{center}
\end{figure}

\begin{figure}[htb]
\begin{center}
\mbox{
\epsfysize4cm
\epsffile{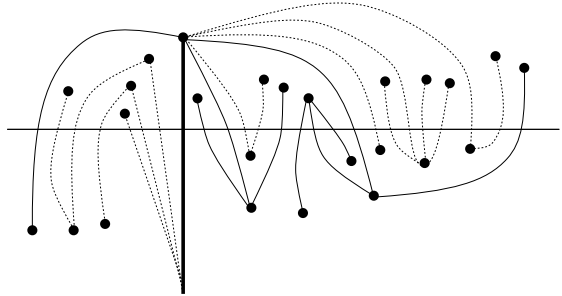}}
\caption{Typical diagram contained in $W(x)$. The root link of the
tree is drawn thicker. The links which occur in a $W$-tree, but not
a $V$-tree are drawn dotted.}
\label{fig:wcomplex}
\end{center}
\end{figure}

The lower bound can be improved further by including more foldings
than exist in the generating function for $V(x)$.
If $V(x)$ is replaced by $W(x)$, whose generating function is given by
the relation in
fig.~\ref{fig:wgen}, then an improved generating function is
\be
\Theta_{imp.}(x)=\frac{x}{1-2W},
\ee
where
\be
\label{eqn:wgen}
W(x)=\frac{x}{(1-\frac{W}{1-W})^2} = x \frac{(1-W)^2}{(1-2W)^2}.
\ee
The generating function for $W(x)$ includes all the trees contained in
$V(x)$, but also trees in which there are arbitrary numbers of
$W$-trees (emerging from the root) interleaved with the branches
connected to the vertex. Replacing $V$ by $W$ in (\ref{eqn:vgen}) and
then changing $W$ to $W/(1-W)$ on the right hand side gives (\ref{eqn:wgen}).
This last replacement takes into account the fact that each branch in
fig.~\ref{fig:vgen} now has an arbitrary number of trees inserted
between it and the previous branch or trunk. Figure~\ref{fig:wcomplex}
shows a typical diagram included in $W(x)$.

Equation (\ref{eqn:wgen}) gives
\be
\frac{dW}{dx}=\frac{(1-W)^2}{(1-8W+12W^2-2x(W-1))}.
\ee
It is the divergence of this derivative which defines the critical
value of $x$,
\be
x_c = \frac{1}{8} \left(71-17 \sqrt{17}\right)
\ee
and hence the lower bound on $R$ is
\be
R \ge {x_c}^{-\half} \approx 2.970.
\ee
This is a significant improvement over the lower bound $R \ge 2$ that we
had initially. It is worth noting that computer enumerations of the
number of semi-meanders yield an estimated value of $R\approx 3.50$
(see reference~\cite{FGG96}). A power series expansion of $\Theta_{imp.}(x)$
gives
\be
\Theta_{imp.}(x) = x + 2 x^2 + 8 x^3 + 42 x^4 + \underline{252} x^5 + 
\underline{1636} x^6 + \cdots,
\ee
which should be compared with the correct series for meander diagrams
\be
M(x)=  x + 2 x^2 + 8 x^3 + 42 x^4 + 262 x^5 + 1828 x^6 + \cdots.
\ee
The two series agree up to and including terms of order $x^4$, after which
we see that $\Theta_{imp.}(x)$ is, as expected, undercounting the number
of  meander graphs (the relevant terms in $\Theta_{imp.}$ are underlined). 
In principle one could improve the generating
functions $\Theta_{imp.}(x)$ and $W(x)$ in order to raise the lower
bound further,
however it appears to require increasing amounts of effort for diminishing
returns.

\section{Diagrammatic method}
\label{sec:diagram}

In this section, we will develop a diagrammatic method for generating
meander and semi-meander diagrams. By truncating some of the
equations, we will derive approximate solutions for the critical
behaviour of $M(x,m)$ and $S(c,m)$, which can be compared with results
from computer enumerations carried out by other authors.

\subsection{$u$-diagrams}
\begin{figure}[bt]
\begin{center}
\mbox{
\epsfysize1.8cm
\epsffile{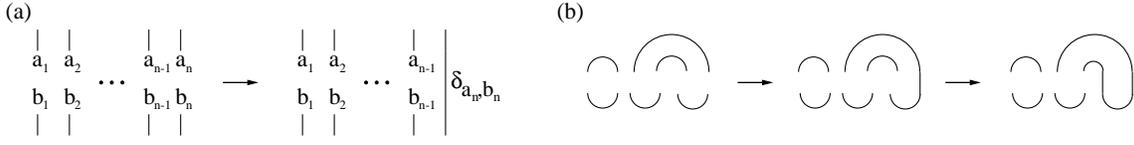}}
\caption{(a) Applying equation (\ref{eqn:therule}) to $u_{a_1}\cdots
u_{a_n} u_{b_n}^\dagger \!\!\! \cdots u_{b_1}^\dagger$.
\ \ (b) Applying equation (\ref{eqn:therule})
twice to a term in $GG^\dagger$.}
\label{fig:glueit}
\end{center}
\end{figure}

Following reference~\cite{MkCh96}, let us introduce a set of
non-commuting variables $\{ u_a, u_a^\dagger\}$ with $a=1, \cdots, m$, which
obey the relation
\be
\label{eqn:therule}
u_a u_b^\dagger = \delta_{ab}.
\ee
We will consider $u_a$, $u_a^\dagger$ as annihilation and creation
operators in some Hilbert space, with a vacuum $|\Omega\rangle$,
such that
\be
\label{eqn:vacuum}
u_a|\Omega\rangle =0 , \mgap \langle \Omega| u_b^\dagger =0, \mgap
\langle \Omega| \Omega \rangle =1.
\ee
Now define $G$ and $G^\dagger$ by
\begin{eqnarray}
\label{eqn:grule}
G&=&1+c \ G u_a G u_a \\ \nonumber
G^\dagger &=&1 + c \ u_a^\dagger G^\dagger u_a^\dagger G^\dagger,
\end{eqnarray}
where we are summing over the repeated index $a$ ($a=1, \cdots, m$).
Diagrammatically we can write the first equation as
\setlength{\unitlength}{0.1mm}
\begin{picture}(305,50)
\thicklines
\put(255,5){\oval(80,80)[t]}
\put(0,0){\makebox(0,0)[bl]{$G=1 \ + \ G \ \ \ G \ $}}
\put(205,-15){a}
\put(285,-15){a}
\end{picture},
where a line emerging from an '$a$', that is written below the equation,
represents $u_a$. 
The fact that the two lines are joined into an
arch indicates that we are summing over the repeated index '$a$'.
Since '$a$' is just a dummy index it can be omitted.
It is to be understood that each arch has a factor of $c$ associated with it.
Expanding this equation gives
\be
G^{\phantom{\dagger}}=1 
\plus \one \plus \one  \one \plus \two \plus \one \one \one
\plus \one \two \plus \two \one \plus \twoinone \plus \three \plus
\cdots,
\ee
that is, $G= 1+ u_a u_a + u_a u_a u_b u_b + u_a u_b u_b u_a
+ \cdots$.
Thus $G$ generates arch configurations, however unlike $A$ in equation
(\ref{eqn:arelation}) $G$ keeps track of where the ends of the arches
are positioned. The generating function $G^\dagger$ generates arch
configurations in terms of $u_a^\dagger$, diagrammatically one can
consider these as inverted arch diagrams. So 
\be
G^\dagger=1
\plus \bone \plus \bone  \bone \plus \btwo \plus \bone \bone \bone
\plus \bone \btwo \plus \btwo \bone \plus \btwoinone \plus \bthree \plus
\cdots. 
\ee
The dagger operator acting on a diagram can be thought of as
reflecting the diagram, for example,
\be
\left[ \one \two \right]^\dagger = \bone \btwo.
\ee
Now consider the expression $G G^\dagger$, this contains terms of the
form $u_{a_1} u_{a_2} \!\!\cdots u_{a_n} u_{b_n}^\dagger u_{b_{n-1}}^\dagger
\!\!\!\!\!\!
\cdots u_{b_1}^\dagger$. When this term is drawn as a diagram the
inverted arches can be drawn under the upright arches to make the
connection with meander diagrams clear. 
One can apply equation (\ref{eqn:therule})
repeatedly to eliminate $u$ and
$u^\dagger$ variables. In terms of diagrams this corresponds to gluing
together a leg on the upright arch diagram to one on the inverted diagram
(see fig.~\ref{fig:glueit}). If the
number of $u$ variables is equal to the number of $u^\dagger$
variables then we will end up with a (possibly disconnected)
meander diagram. However, if the numbers are not equal then there will
be spare $u$ or $u^\dagger$ left over. Taking the vacuum expectation
value will kill these extra terms and hence $\langle G G^\dagger
\rangle$ generates all the meander diagrams, since any meander can be
made by gluing together two arch diagrams.
Each closed loop in the
meander diagram will give a factor of $\delta_{a a}= m$. Thus the
correct weightings are generated and hence
\be
\langle G G^\dagger\rangle= 1 + M(x,m).
\ee

\subsection{$X$-diagrams}
It is convenient now to introduce a new set of variables, $\{X_a,
X_a^\dagger\}$, defined by
\be
X_a= \frac{G u_a}{\sqrt{m}}, \mgap 
X_a^\dagger =\frac{u_a^\dagger G^\dagger}{\sqrt{m}}.
\ee
So that
\be
\xx \equiv \sum_{a=1}^m X_a X_a^\dagger = G G^\dagger,
\ee
\be
\left\langle\xx \right\rangle=1+M(x,m).
\ee
The equations for $X_a$ equivalent to (\ref{eqn:therule}) and 
(\ref{eqn:vacuum}) are
\be
\label{eqn:xtherule}
X_a X_b^\dagger = \frac{1}{m} \delta_{ab} \xx,
\ee
\be
\label{eqn:xvacuum}
X_a|\Omega\rangle =0 , \mgap \langle \Omega| X_b^\dagger =0, \mgap
\langle \Omega| \Omega \rangle=1.
\ee
From (\ref{eqn:grule}) we have
\begin{eqnarray}
\xx& =& \left(1+c m \xa \xa \right) \left(1+c m \xbd \xbd \right)\\
\label{eqn:master}
&=& 1 +c m \left(\xa\xa +\xad\xad \right) + c^2 m \xa \xx \xad,
\end{eqnarray}
where repeated indices are summed over.
In a similar fashion to the $u,u^\dagger$ diagrams, this can be
written diagrammatically, in this case as
\be
\label{eqn:xgendiag}
\xsplit = 1 + cm\left( \one \ \xline + \xline  \ \one \right)+ c^2 m \ 
\xsplittwo,
\ee
where the ends corresponding to non-daggered operators are on the
left of the vertical line and those for the daggered operators are on
the right. Two ends are joined into an arch whenever the corresponding
operators share a summed over index. The vertical line is only present
in order to indicate the boundary between non-daggered and daggered
operators.
 Equation (\ref{eqn:xtherule}) implies that
\be
 \xa  \xa \xbd  \xbd  =
\frac{1}{m} \left(  \xa  \xc \xcd  \xad  \right),
\ee
 which was used above, and
\be
 \xb  \xa  \xa \xbd  =
\frac{1}{m} \left(  \xa  \xa  \xc \xcd  \right),
\ee
which can be drawn respectively as
\be
\one \ \xline \ \one = \frac{1}{m} \xsplittwo
\ee
and
\be
\xsplittwoleft = \frac{1}{m} \one \xsplit \ .
\ee
In principle one could repeatedly substitute equation
(\ref{eqn:master}) into itself in order to expand $\xx$ in terms of
$\xa$ and $\xad$. However this rapidly becomes tedious and it is
necessary to find a simpler way of calculating $\xx$.
Let us define $\{ A_n\}$ by
\be
A_0=1, \mgap A_1=\xsplit, \mgap A_2=\xsplittwo, \mgap
A_3=\xsplitthree, \ \ \cdots \ 
\ee
and $\{ Y_n\}$ by
\be
\label{eqn:ydef}
Y_1= c m \left( \one \ \xline + \xline \ \one   \right), \mgap
\begin{picture}(220,60)
\put(160,0){\oval(120,100)[t]}
\put(0,0){$Y_n= \ \ Y_{n-1} \ \ $}
\end{picture} \ .
\ee
For later use we will define the following notation, that
\toright is an operator, which multiplies the expression following it
by $\xa \xa$ (that is, \one ) on the left-hand side, 
whereas \toleft multiplies the expressing following it by
$\xad \xad$ on the right-hand side. Thus,
\be
Y_1= c m \left( \one \ \xline + \xline \ \one   \right) =
     c m \left( \toright + \toleft \right) \xline \ 
\ee
and
\be
Y_2= c \left( \one \xsplit + \xsplit \one \right) = 
     c \left( \toright + \toleft \right)  \xsplit \ .
\ee

Now equation (\ref{eqn:xgendiag}) can also be written as
\be
A_1=1+Y_1 + \alpha A_2 ,
\ee
where $\alpha \equiv c^2 m = x m$.
This equation generalizes to
\be
\label{eqn:aygen}
A_n= A_{n-1} + Y_n + \alpha A_{n+1}.
\ee
Note that we wish to evaluate $\lang A_1 \rang$ since
\be
\lang A_1 \rang = \lang \xsplit \rang
= \lang \xx \rang = \lang G G^\dagger \rang = 1+M(x,m).
\ee
The next step is to rewrite $A_1$ in terms of $\{ Y_n \}$ and $\alpha$.
We rewrite (\ref{eqn:aygen}) as
\be
\label{eqn:anbargen}
0= \alpha \left( A_{n+1} - f A_n \right) - \frac{1}{f}
\left( A_n - f A_{n-1} \right) +Y_n ,
\ee
where $f(\alpha)$ is given by 
\be
\alpha f + \frac{1}{f} =1, \mgap f(\alpha)= 
\frac{1}{2 \alpha} \left(1- \sqrt{1-4 \alpha} \right)
\ee
and define $\overline{A}_n$ by
\be
\overline{A}_n=A_n - f A_{n-1}.
\ee
Then from (\ref{eqn:anbargen})
\be
\overline{A}_n=  f Y_n +\alpha f \overline{A}_{n+1}.
\ee
This yields immediately
\be
\label{eqn:useful}
A_1 =f +  \overline{A}_1 =
 f.\left(1 + Y_1 + (\alpha f) Y_2 + (\alpha f)^2 Y_3 + \cdots\right).
\ee
The first term in $\lang A_1 \rang$ is $f(\alpha)$, which generates
the Catalan numbers (see equation (\ref{eqn:adef})). These correspond
to the set of numbers $M_n^{(n)}$ in $M(x,m)$, that is, the leading
diagonal of table~\ref{tab:numbers}. 

\subsection{Calculation of $\lang Y_n \rang $}
In order to calculate $\lang A_1
\rang$, or at least an approximation to it, we need to consider how
to calculate $\lang Y_n \rang$.
It is worth noting that $\lang Y_1 \rang=0$ from (\ref{eqn:xvacuum}).

Now consider $Y_2$,
\be
Y_2=c \tofact A_1 = c f \tofact \left( 1 + Y_1 + (\alpha f) Y_2 +
(\alpha f)^2 Y_3 + \cdots \right),
\ee
so that
\be
Y_2 = \left(1- c \alpha f^2 \tofact \right)^{-1} \left[ cf \tofact
(1+Y_1) + O(Y_3) \right].
\ee
So far all the equations have been exact, but from now on in this
expression we shall
ignore the terms $Y_n$ for which $n \ge 3 $. This will make the
calculation more tractable, but will cause us to miss out some of the
meander diagrams in our generating functions. That is, the equation
for $\lang A_1 \rang$ will contain only a subset of the meanders. Now,
\be
Y_2 = cf  \sum_{n=0}^{\infty}  \left(c \alpha f^2 \right)^n
\tofact^{n+1} \left[ 1+ cm \tofact \right].\ \xlinelong \ + O(Y_3)
\ee
and rearranging this gives
\be
Y_2 = \frac{fY_1}{m} + x m f \left(1+ x f^2 \right) \sum_{n=1}^\infty
\left(c \alpha f^2 \right)^{n-1} \tofact^{n+1}. \ \xlinelong + O(Y_3).
\ee
This summation will be truncated so as to keep terms up to and
including $n=2$, again this will cause us to lose some diagrams,
\begin{eqnarray}
Y_2 = \frac{fY_1}{m} &+& x m f \left(1+ x f^2 \right) 
\left[ \tofact^2 + \left(c \alpha f^2 \right) \tofact^3 \right] . \
\xlinelong \nonumber \\
&+& O(Y_3) + O\left(\tofact^4\right). \label{eqn:whytwo}
\end{eqnarray}
From now on the omitted terms will just be indicated by three dots,
however it should be understood that this means terms of the form
indicated in the second line of the above equation.
To simplify this equation, consider that,
\begin{eqnarray}
\tofact^2 \dotline &=& \twofact
 \dotline + 2 \Bigl( \toright \toleft \Bigr) \dotline \nonumber \\
&=& \twofact \dotline + \frac{2}{m} \xsplittwo
\end{eqnarray}
and similarly
\be
\tofact^3 \dotline = \threefact \dotline + \frac{3}{m} \tofact
\underbrace{\xsplittwo}_{A_2} .
\ee
Now from (\ref{eqn:useful}), $A_2 = f \left( A_1 + Y_2 \right) + O(Y_3)$
so that
\begin{eqnarray}
\!\!\! \tofact^3 \dotline &=& \threefact \dotline + \frac{3f}{m} \tofact
\left(A_1 + Y_2 \right) + \cdots \nonumber \\
&=&\threefact \dotline + \frac{3f}{mc} Y_2 + \frac{3fc}{m} \tofact^2
A_1 + \cdots, \label{eqn:etcetera} 
\end{eqnarray}
where we can use (\ref{eqn:useful}) to eliminate $A_1$. Basically one
can repeat this process of substituting equations into
(\ref{eqn:etcetera}) to gradually increase the coefficient in front of
$Y_2$, which has the effect of improving the accuracy of the final
result. The actual equation that is finally used is somewhat arbitrary,
but eventually one gains an expression such as
\begin{eqnarray}
\tofact^3 \dotline &=& \left( 1 - 3 x f^2 - 3 x^2 f^4\right) ^{-1}
\Biggl[\threefact \dotline  \nonumber \\
&+& \frac{3f}{mc}Y_2 +\frac{3 f^2 c}{m}
\tofact^2 \dotline \ \Biggr] + \cdots.
\end{eqnarray}
Substituting this into (\ref{eqn:whytwo}) gives, after a bit of
manipulation,
\be
\label{eqn:ytwo}
Y_2 = \frac{1}{D}  \left( E.Y_1 +F.A_2 + \frac{Fm}{2} \twofact
\dotline + J \threefact \dotline \ \right) + \cdots ,
\ee
where
\begin{eqnarray}
D &\equiv& 1-3 x f^2 - 3 x^2 f^4 \left(1+m \left( 1+x f^2 \right)
\right), \nonumber \\
E &\equiv& \frac{f}{m} \left( 1 - 3 x f^2 - 3 x^2 f^4 \right), \nonumber \\
F &\equiv& 2 x f \left(1+x f^2 \right) \left(1 - 3 x f^2 \right), \nonumber \\
J &\equiv& x^2 m^2 f^3 c \left( 1 + x f^2 \right). \label{eqn:defdtog} 
\end{eqnarray}
This gives us 
\be
\lang Y_2 \rang = \frac{F}{D} \lang A_2 \rang + \cdots = a' \lang A_2
\rang + \cdots,
\ee
where we define $a'=F/D$.
To calculate the critical behaviour it is necessary to have a formula
for $\lang Y_3 \rang$. This can be derived from (\ref{eqn:ytwo}) using
(\ref{eqn:ydef}), that is,
\be
\label{eqn:ythree}
Y_3 =  \frac{1}{D}  \left( E.Y_2 +F.A_3 + \frac{Fm}{2} \twounder
 + J \threeunder \right) + \cdots .
\ee
Now,
\be
\twounderplain = \frac{1}{m} \left( \two \xsplit + \xsplit \two \right),
\ee
where using (\ref{eqn:useful}) to substitute for $\xsplit$, and
keeping only the relevant terms, we have
\begin{eqnarray}
\twounderplain &=&  cf \left( \two \ \xline \ \one + \one \ \xline \
\two \right) +\cdots \nonumber\\
&=& \frac{cf}{m} \left( \xisleft + \xisright \right) + \cdots
= \frac{Y_3f}{m} +\cdots.
\end{eqnarray}
In a similar fashion, we can extract a term proportional to $Y_3$ by
the following manipulation,
\begin{eqnarray}
\threeunder \!\!\!\! &=& \!\!\! \frac{1}{m} \left( \twoinone \xsplit + \xsplit
\twoinone \right)
= cf \left( \twoinone \ \xlinelong \ \one + \one \ \xlinelong \ \twoinone
 \right) + \cdots \nonumber \\
&=&\!\!\! \frac{cf}{m} \left( \twosplitplain \right) + \cdots
\nonumber \\
&=& \!\!\! \frac{c f^2}{m} \twounder + \cdots
= \frac{c f^3}{m^2} Y_3 + \cdots.
\end{eqnarray}
Substituting this into equation (\ref{eqn:ythree}) gives 
\be
Y_3=\frac{1}{D} \left(E.Y_2 + F.A_3 + \half Ff Y_3 +
\frac{J c f^3}{m^2} Y_3 \right) + \cdots.
\ee
Rearranging this gives us,
\be
Y_3 = \frac{F}{H}  A_3  + \frac{E}{H}  Y_2
+ \cdots =a A_3  + b  A_2  + \cdots,
\ee
where we have defined
\be
\label{eqn:defhab}
H \equiv D- \half Ff - \frac{J c f^3}{m^2}, \mgap
a \equiv \frac{F}{H}, \mgap b \equiv a \frac{E}{D}.
\ee
This relation generalizes to
\be
Y_n=a A_n + b A_{n-1} + \cdots, \mgap {\rm for \ } n\ge 3.
\ee
\subsection{Meander generating function}
Now
\be
\label{eqn:anrel}
A_n=A_{n-1} + \alpha A_{n+1} + Y_n,
\ee
so that for $n \ge 3$,
\be
0=\alpha A_{n+1} + (a-1) A_n + (b+1) A_{n-1} + \cdots.
\ee
Let us define an operator $\oh$, which adds an outer arch to the
diagram that it is operating on, that is, $\oh Y= \xa Y \xad$. Then
since $\oh A_n=A_{n+1} $,
\be
0=\alpha \oh^2 A_2 + (a-1)\oh A_2 + (b+1) A_2 + \cdots
\ee
and hence multiplying by $\exp(y\oh)$ gives
\be
0=\left(\alpha \left(\ddy\right)^2  + (a-1)\ddy  + (b+1)\right) 
\lang \exp(y \oh) A_2 \rang + \cdots.
\ee
Keeping only the explicitly written term, the solution is
\be
\lang \exp(y \oh) A_2 \rang = C e^{ky},
\ee
where
\be
\label{eqn:defk}
k= \frac{1}{2\alpha} \left( 1-a - \sqrt{(1-a)^2 - 4 \alpha (b+1)}\right).
\ee
So that $\lang A_n \rang = C k^{n-2}$ for $n\ge 2$. Using the relation
(\ref{eqn:anrel}) for $n=1,2$ gives us the approximate result
\be
\lang A_1 \rang = 1+ \alpha C, \mgap
C =\frac{1}{1-a' - \alpha - \alpha k}.
\ee
Expanding this using equations (\ref{eqn:defdtog}), (\ref{eqn:defhab})
and (\ref{eqn:defk}) gives
\begin{eqnarray}
\lang A_1 \rang &=&  1+mx + (2 m + 2 m^2 ) x^2 + (8 m+12 m^2+5
m^3)x^3 \nonumber \\
&\ & + ({\underline{38}} m+84 m^2+56 m^3+14 m^4) x^4 + O(x^5) +
{\rm missing\ diagrams}.
\end{eqnarray}
Comparing with table~\ref{tab:numbers}, we find that the approximate
formula generates the correct coefficients up to $x^4$, except for the
coefficient of $mx^4$, which is lower than it should be.
Expanding the series further one sees that more and more
of the coefficients are below the correct value, that is, as expected
we are undercounting the number of diagrams as a result of the terms
that were dropped during the calculation.

Now the singular behaviour of $\lang A_1 \rang $ is given by the
singularity in $k$ caused by the vanishing of the square root. This
gives a relation that determines approximately $x_c$, the critical
value of $x$, as a function of $m$. The quantity $x_c(m)$ has been
calculated numerically using the equations given and the results are
plotted in figure~\ref{fig:graph}.

\subsection{Semi-meanders}
So far we have only discussed how to apply the $X$-diagrams to the
problem of enumerating meanders, but in fact they can also be used to
calculate semi-meanders.

\begin{figure}[htb]
\begin{center}
\mbox{
\epsfysize3cm
\epsffile{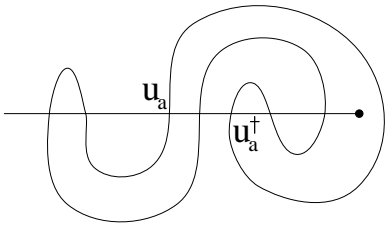}
}
\caption{A typical semi-meander with winding number of one.}
\label{fig:semiug}
\end{center}
\end{figure}

Suppose we consider a semi-meander with winding number equal to one
(fig.~\ref{fig:semiug}). Note that the winding number (denoted by $w$)
is the minimum number of bridges that would need to be added if we
were to extend the river to infinity on the righthand side. Meanders
are just semi-meanders with winding number zero and are generated by
$\lang G G^\dagger \rang$. Semi-meanders with $w=1$ are generated by
the expression $c\lang G u_a G G^\dagger u_a^\dagger G^\dagger \rang$.
That is, the upper part of the diagram consists of two arch
configurations separated by a $u_a$, which is linked to a
$u_a^\dagger$ below, that separates two inverted arch configurations.
More generally a semi-meander with winding number $w$ is generated by
$c^w \lang G u_{a_1} G \cdots u_{a_w} G G^\dagger u^\dagger_{a_w}
\cdots G^\dagger u_{a_1}^\dagger G^\dagger  \rang $.
So that the generating function for semi-meanders is
\be
S(c,m)=\lang \xa \xad \rang + cm \lang \xa \xb \xbd \xad \rang + 
c^2 m^2 \lang \xa \xb \xc \xcd \xbd \xad \rang+\cdots,
\ee
which can be rewritten as
\be
\label{eqn:sresult}
S(c,m)= \sum_{w=0}^\infty \left(cm \right)^{w} \lang A_{w+1} \rang .
\ee
Using our approximate formula for $\lang A_n \rang $ this gives
\be
S(c,m)=1+ C m \left( x + \frac{c}{1-mck} \right).
\ee
A power series expansion of this expression gives
\begin{eqnarray}
S(c,m) &\!\! = \!\!& 
1 + m c + (m + m^2 ) c^2 + (2m+ 2m^2 +m^3) c^3 + (4 m + 6 m^2
+ 3 m^3 + m^4) c^4 \nonumber \\
& \ & + (\underline{8}m +16 m^2+11 m^3+4 m^4 + m^5) c^5 +
O(c^6)+ {\rm missing\ diagrams} ,
\end{eqnarray}
which is correct up to $O(c^5)$ except for the coefficient of $m c^5$.
As expected this formula undercounts the number of diagrams.

\begin{figure}[phtb]
\begin{center}
\mbox{
\epsffile{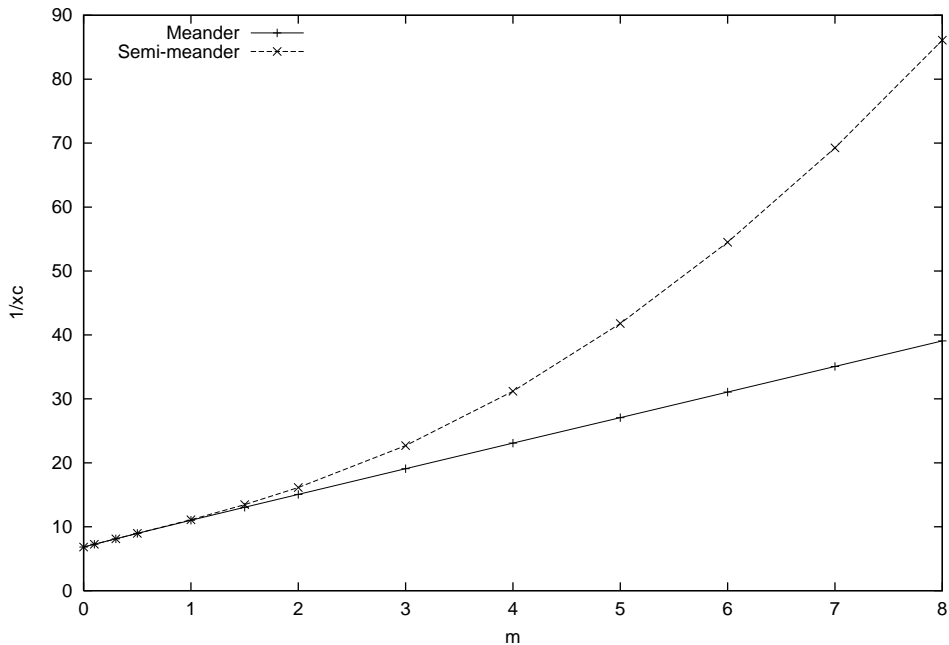}
}
\caption{Plot of $1/x_c$ as a function of $m$ for meanders and
semi-meanders using the approximate formulae.}
\label{fig:graph}
\end{center}
\end{figure}

The critical behaviour for semi-meanders is caused by the vanishing of
the quantity $(1-mck)$ for large $m$, and by the vanishing of the
square root in the formula for $k$ at small $m$ (that is, $m \ltsim 0.6$). 
The function $x_c(m)$ for semi-meanders has
been calculated numerically and is displayed in
figure~\ref{fig:graph}.

\begin{figure}[phtb]
\begin{center}
\mbox{
\epsffile{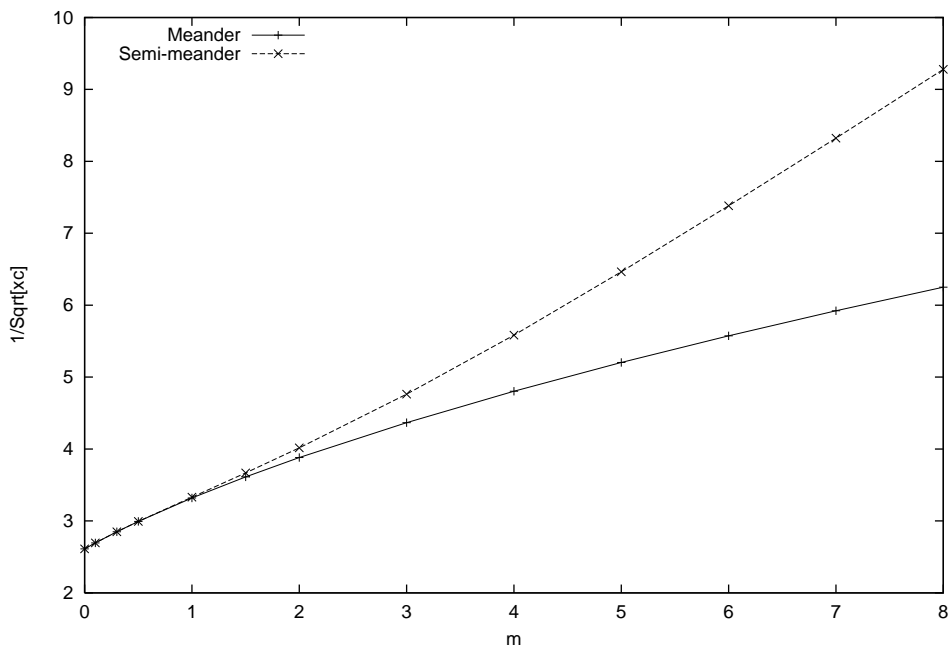}
}
\caption{Plot of $R(m)=1/\sqrt{x_c}$ for meanders
and $\rb (m)=1/\sqrt{x_c}$ for semi-meanders, using the approximate
formulae.}
\label{fig:graph2}
\end{center}
\end{figure}

\subsection{Discussion}
\label{sec:disc}
The graph (fig.~\ref{fig:graph}), 
which plots the approximate value of $1/x_c$ as a function
of $m$ for both meanders and semi-meanders, shows a number of
interesting features. First of all it should be noted that the missing
diagrams in the power series expansions for $M(x,m)$ and $S(c,m)$
tend to be those with lower powers of $m$, whereas the coefficients
of the highest powers of $m$, ($m^n x^n$ for meanders and $m^n c^n$ 
for semi-meanders,) are exact. Thus one would expect our approximation
to be exact in the $m\to \infty$ limit and least accurate in the $m \to
0$ limit. Unfortunately, the most interesting behaviour occurs for
small values of $m$. 

In reference~\cite{FGG96} the authors perform a
computer enumeration of the semi-meander diagrams. They conclude that in
the function $S(c,m)$, there is a phase transition at $m\approx 2$, between a
meander-like phase (with irrelevant winding number)
and a phase in which winding
is relevant. Figure~\ref{fig:graph} shows indications
of such a phase transition developing at around $m\approx 0.6$. Below this
value of $m$, $x_c$ for semi-meanders is equal to that for
meanders. Above $m \approx 0.6$ the two curves split apart. 
This is caused by a change in the form of the critical behaviour for
the semi-meanders, as indicated in the previous section. In the
next section we will examine the behaviour of the winding number for
semi-meanders as a function of $m$, and this also gives evidence of a phase
transition at $m \approx 0.6$.
One would expect that, as our approximation is improved by including
more of the missing diagrams, the location 
of the phase transition would become clearer and be in closer agreement
with ref.~\cite{FGG96}.

Figure~\ref{fig:graph2} shows our approximation to
$1/\sqrt{x_c}$ as a function of $m$. This can be directly compared
with fig.~9 of ref.~\cite{FGG96}, which calculates the same quantities,
but using a different method.
The two graphs are very similar, although it is immediately apparent
that the undercounting of diagrams has pushed the curves on our graph
lower than they should be. In particular, that paper derived an exact
value for $R(m)=1/\sqrt{x_c}$ (for meanders) and $\rb (m)$
similarly defined for semi-meanders. The result they gave was $R(1)=
\rb (1)=4$, showing that our approximate value of
$R(1)=\rb (1)=3.3$ is too low.
Even so the general form of our graphs is the same as that derived
in ref.~\cite{FGG96}, lending support to the results
in that paper. 

In the limit $m \to 0$ our approximation gives $R(0)=\rb
(0)=2.6$. Again this is too low as we have already proved in
section~\ref{sec:further} that $R(0) \ge 2.970$. Reference~\cite{FGG96}
on the other hand has $R(0)=\rb (0)=3.5$, which is comfortably within
the known bounds.

It should perhaps be noted that the paper~\cite{FGG96} also showed that
for $m\to \infty$ one has $R \sim \sqrt{m}$ and $\rb \sim m$. This
behaviour can clearly be seen in the
linearity of the lines for meanders (in fig.~\ref{fig:graph})
and semi-meanders (in fig.~\ref{fig:graph2}) at large $m$.

\subsection{Winding number}
\label{sec:wind}
\begin{figure}[htb]
\begin{center}
\mbox{
\epsffile{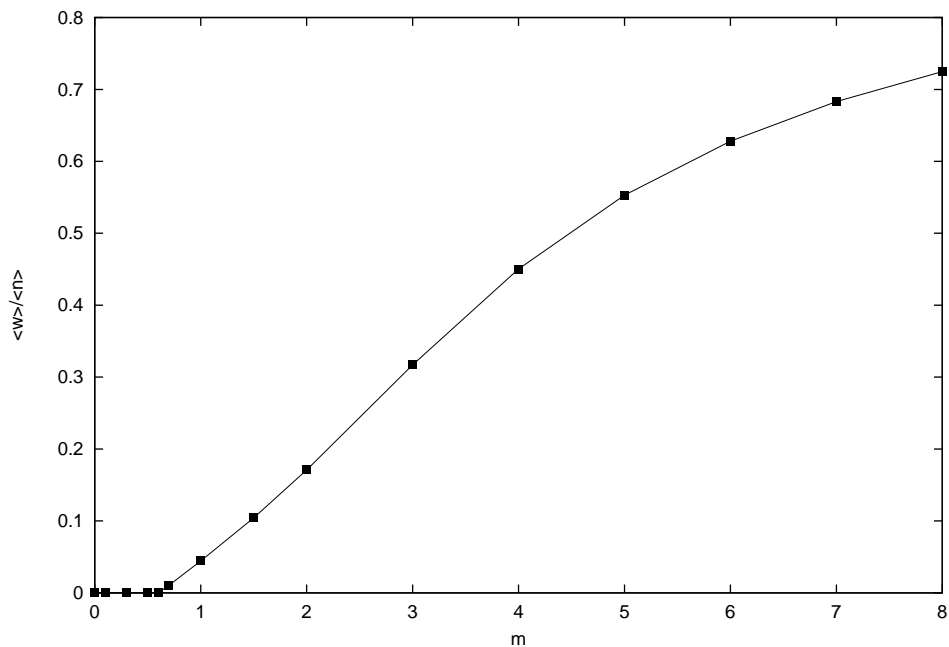}
}
\caption{Graph of $\lang w\rang / \lang n \rang$ plotted as a function
of $m$, showing evidence of a phase transition.}
\label{fig:wind}
\end{center}
\end{figure}

The phase transition for the semi-meander diagrams is believed to be a
winding transition with an average winding per bridge equal to zero
in one phase and non-zero in the
other. This suggests that we should examine more closely the winding
numbers of diagrams in the semi-meander generating function $S(c,m)$.
From (\ref{eqn:sresult}) we see that
\be
\lang w \rang = \frac{1}{S(c,m)} \sum_{w=0}^\infty  (cm)^w w 
\underbrace{\lang A_{w+1} \rang}_{C k^{w-1}} = \frac{C m c}{S (1-mck)^2}.
\ee
Now from (\ref{eqn:defsemi}) the average number of bridges $\lang n\rang$ is
given by
\be
\lang n \rang = \frac{c}{S} \frac{\partial S}{\partial c}.
\ee
So that
\be
\label{eqn:windings}
\frac{\lang w \rang}{\lang n \rang} = \frac{C m}{(1-mck)^2} \left(
\frac{\partial S}{\partial c} \right)^{-1},
\ee
which gives a measure of the average number of windings per bridge;
we expect that $0 \le \lang w \rang / \lang n \rang \le 1$. Using a
symbolic algebra program to calculate (\ref{eqn:windings}) and evaluating for
different values of m (using our previously calculated values of
$x_c(m)$)  gives us the graph shown in
fig.~\ref{fig:wind}. This graph shows clear evidence of a phase
transition occurring at about $m \approx 0.6$. As discussed in the
previous section we are using an approximate generating function,
which is known to be inaccurate for small values of $m$, so this
result should be treated cautiously. However it is at least
encouraging to see such clear signs of a phase transition, and one
could hope that further effort to improve the $m \to 0$ behaviour of
the generating functions would increase the accuracy of this estimate.

\section{Conclusion}
\label{sec:concl}
In this paper we have shown how one can tackle the meander problem
using diagrammatical techniques. In section~\ref{sec:three} the
problem of counting connected meanders was converted to one of
enumerating folded trees. Using this connection it was shown that
\be
2.970 \le R \le 4,
\ee
where $R \equiv R(0)$ is defined by (\ref{eqn:masymp}) and
gives a measure of the
exponential growth of connected meanders as the number of bridges is
increased. This is an improvement over the usual bounds of $2 \le R
\le 4$, 
and is consistent with the numerical estimate~\cite{FGG96} of $R
\approx 3.5$. 
It may be possible to improve this lower bound, by modifying
the method to increase the number of folded configurations that are
included. Significant improvements seemed to be quite difficult to
achieve, but the idea may be worth further study.

In section~\ref{sec:diagram} we used a non-commutative representation
of the meander problem to calculate approximate formulae for the
generating functions of both meanders and semi-meanders.
The approximations were necessary in order to make the equations
tractable and resulted in the loss of some diagrams, which should have
been present in the generating functions. Nevertheless we have managed
to extract graphs showing the behaviour of $x_c$ for the two models,
which were qualitatively the same as those generated using a different
method~\cite{FGG96}. In particular we have shown that there is
a region over which
$\rb (m)$, for the semi-meanders, is equal to $R(m)$, for
the meanders, giving an indication of the existence of a phase
transition. Examination of the average winding number per bridge shows
much clearer evidence of the phase transition and yields an estimated
critical value of $m_{c} \approx 0.6$. Comparison with
ref.~\cite{FGG96}, which has $m_{c} \approx 2$, suggests that our
estimate is too low.
It would seem to be worthwhile to attempt to improve the approximations
used, in order to narrow down the exact location of this phase
transition. Many questions remain unanswered concerning the meander
problem and the critical behaviour of the semi-meander generating
function, and these must be left for future investigation.

\subsection*{Acknowledgements}
MGH would like to acknowledge the support of the European Union
through their TMR Programme, and to thank Charlotte Kristjansen and
Yuri Makeenko for interesting discussions.

\end{document}